\documentstyle[preprint,aps,epsf]{revtex}

\def\NPB#1#2#3{{\em Nucl. Phys.} {\bf B#1} (19#2) #3}
\def\PLB#1#2#3{{\em Phys. Lett.} {\bf B#1} (19#2) #3}

\def\PRD#1#2#3{{\em Phys. Rev.} {\bf D#1} (19#2) #3}
\def\PRL#1#2#3{{\em Phys. Rev. Lett.} {\bf#1} (19#2) #3}

\def\ZPC#1#2#3{{\em Zeit. f\"ur Physik} {\bf C#1} (19#2) #3}

\def\NC#1#2#3{{\em Nuovo Cim.} {\bf A#1} (19#2) #3}

\begin{document}
\newcommand{\cross}{\mbox{$\rlap{\kern0.125em/}$}}
\draft
%
\title{Tevatron Potential for 
Technicolor Search with Prompt Photons.}
%
\author{Alexander Belyaev $^{1,2}$, Rogerio Rosenfeld$^1$
and Alfonso R. Zerwekh$^1$ }
\address{ 
$^1$ {\it Instituto de F\'\i sica Te\' orica, Universidade 
Estadual Paulista},\\
Rua Pamplona 145, 01405-900 - S\~ao Paulo, S.P., Brazil\\
$^2$ {\it Skobeltsyn Institute for Nuclear Physics,
Moscow State University, \\ 119 899, Moscow, Russian Federation}
}
\maketitle
\widetext
\vspace*{0cm}
\begin{abstract}
We perform a detailed study of the process of single color
octet isoscalar $\eta_T$ production at the Tevatron with  $\eta_T \rightarrow
\gamma + g$ decay signature, including a
complete simulation of signal and background processes.
We determined a set of optimal cuts from an analysis of the various kinematical 
distributions for the signal and backgrounds. 
As a result we show the exclusion and discovery limits on the $\eta_T$ mass
which could be established at the Tevatron for some technicolor models.

\end{abstract}
\vspace*{0cm}
\vspace*{0.5cm}
\section{Introduction}

The Standard Model has been extensively tested and it is a very successful
description of the weak interaction phenomenology. Nevertheless, the electroweak
symmetry breaking sector has essentially remained unexplored.  In the Standard
Model, the Higgs boson plays a crucial r\^ole in the symmetry breaking
mechanism. However, the presence of such a  fundamental scalar at the 100 GeV
scale gives rise to some theoretical problems, such as the naturalness of a
light Higgs boson and the triviality of the fundamental Higgs self-interaction.
These problems lead to the conclusion that the Higgs sector of the Standard
Model is in fact a low energy effective description of some new Physics at a
higher energy scale.

Three main avenues for the new Physics have been proposed: low-scale
supersymmetry\cite{SUSY}, large extra dimensions at TeV scale\cite{extradim}
and dynamical symmetry breaking~\cite{Weinberg79,Suskin79}.  A common
prediction of all these extensions of the Standard Model is the appearance of 
new particles in the range of some hundred GeV to a few TeV.

We  focus on the issue of the technicolor models of dynamical
symmetry breaking which predict new particles such as 
Pseudo-Nambu-Goldstone-Bosons (PNGB) and
vector resonances. Many of these models include colored technifermions and
hence some PNGB's can  be color-triplet or even color-octet particles. These
color-octet scalars can be copiously produced at a hadron collider and they
are the subject of this paper.

The main contribution to the color-octet PNGB
masses comes from QCD. If it is assumed that technicolor dynamics scales
from QCD, this contribution is in the range of $200-400$~GeV, but their masses
can be different in models with non-QCD-like dynamics\cite{chivros}. 

Production of  technicolor particles has been studied  at various present
and future colliders such as Tevatron\cite{HCI}, LEP\cite{Lubicz},
NLC\cite{NLC} and the  Muon Collider\cite{TFMC}.  The impact of PNGB on rare K
meson decays induced through the exchange  of color-singlet $\pi_{T}^{\pm}$ and
color-octet $\pi_{T8}^{\pm}$ technipions has been recently 
studied\cite{Klimits} in the context of multiscale technicolor\cite{MTC} where
typical limits of the order of $m_{\pi_{T8}}\geq 250$~GeV were obtained.

Of special interest is the case of the isoscalar color-octet
PNGB, the so--called technieta ($\eta_T$), since it can be
produced via gluon fusion through the heavy quark loop.
In the near future, the upgraded $2$~TeV Tevatron collider
will be the most promising machine for technieta search.
The channel $p\bar{p} \rightarrow \eta_T
\rightarrow t\bar{t}$ was initially  studied by Appelquist and
Triantaphyllou~\cite{ttbar1} in the context of the one family technicolor
model\cite{OFTC}. More recently  Eichten and Lane\cite{ttbar2} have studied the
same channel in the context of walking technicolor. They concluded that
a technieta with mass in the range $M_{\eta_T}=400-500$~GeV  doubles the
top quark production cross section at Tevatron and hence it is excluded in 
this mass range.

In our study we would like to concentrate on the 
search of technieta with mass below the $t\bar{t}$ threshold,
which is not constrained by
the $t\bar{t}$ production process.

The $p\bar{p} \rightarrow \eta_T \rightarrow gg$ and  $p\bar{p}
\rightarrow \eta_T \rightarrow g\gamma$ processes were studied in the 
early eighties by Hayot and Napoly~\cite{eta1} in the framework of the
one family technicolor model. They
showed that, due to the signal-to-background ratio, the gluon-photon channel
is preferable. Nevertheless, theirs results must be taken only 
as qualitative since no complete and detailed analysis were made.

In this letter we perform a
 complete realistic  study  of the 
$p\bar{p} \rightarrow \eta_T \rightarrow \gamma+\mbox{jet} $ 
process in order to understand
Tevatron potentials for $\eta_T$ search with the mass 
below the $t\bar{t}$ threshold.
We consider  three different scenarios: 
the one family model~\cite{OFTC},
top-color assisted technicolor~(TC2)~\cite{TC2} and
multiscale technicolor~\cite{MTC}.

\section{Effective Couplings}

 The color-octet technieta couples to gluons and photons through the
Adler-Bell-Jackiw anomaly\cite{ABJ}. This effective coupling can be written as:

\begin{equation}
A(\eta_T \rightarrow B_1 B_2)=\frac{S_{\eta_T B_1 B_2}}{4\pi^2 \sqrt{2}
  F_Q}\ \ \epsilon_{\mu \nu \alpha \beta}\epsilon_1^{\mu} \epsilon_2^{\nu}
  k_1^{\alpha} k_2^{\beta}   
\end{equation}
\noindent
where $\epsilon_i^{\mu}$ and $k_i^{\mu}$ represents the polarization
and momentum of the vector boson $i$. In our case the factors
$S_{\eta_T B_1 B_2}$ are given by\cite{Ellis} : 

\begin{equation}\label{etagg}
  S_{\eta_{Ta} g_b g_c}=g_s^2d_{abc}N_{TC}
\end{equation}

\noindent
and

\begin{equation}\label{etagfoton}
  S_{\eta_{Ta} g_b \gamma}=\frac{g_s e}{3} \delta_{ab}  N_{TC}
\end{equation}
where $g_s= \sqrt{4\pi \alpha_s}$, $e=\sqrt{4\pi \alpha}$ and $\alpha$ and 
$\alpha_s$ are the electromagnetic and strong
 coupling constant,  and $N_{TC}$
is the number of technicolors (we take $N_{TC}=4$)
 
The technieta coupling to quarks can be written as:

\begin{equation}\label{etaqq}
  A(\eta_T \rightarrow q \bar{q})=\frac{m_q}{F_Q}\bar{u}_q \gamma_5
  \frac{\lambda_a}{2}v_q . 
\end{equation}

With these couplings we can compute the technieta partial widths:

\begin{equation}
  \label{eq:gamma_eta_gg}
  \Gamma(\eta_T \rightarrow gg)= \frac{5\alpha_S^2 N_{TC}^2
  M^3_{\eta_T}}{384\pi^3 F_Q^2}\ \ ,
\end{equation}

\begin{equation}
  \label{eq:gamma_eta_gfoton}
  \Gamma(\eta_T \rightarrow g\gamma)\left(\frac{N_T e g_s}{4\pi
  F_Q}\right)^2 \frac{M_{\eta}^3}{576\pi} 
\end{equation}

\begin{equation}
  \label{eq:gamma_eta_qq}
  \Gamma(\eta_T \rightarrow q\bar{q})= \frac{m^2_q M_{\eta_T} \beta_q}
  {16\pi F_Q^2}
\end{equation}
where 

\begin{equation}
 \beta_q = \sqrt{1-\frac{4m_q^2}{M^2_{\eta_T}}}\ \ , 
\end{equation}
$m_q$ is the quark mass and $M_{\eta_T}$ is the technieta mass.

These expressions were used to calculate the technieta total
width. From equations (\ref{eq:gamma_eta_gg}) and
(\ref{eq:gamma_eta_gfoton}) we can see that:

\begin{equation}
 \frac{\Gamma(\eta_T \rightarrow \gamma g)}{\Gamma(\eta_T \rightarrow
 g g)}=\frac{2\alpha}{15\alpha_s } = 8.7 \times 10^{-3}. 
\end{equation}
Hence, the decay channel  $\eta_T
\rightarrow g \gamma$ is suppressed,  but due to the more manageable background
it is expected to provide a larger statistical significance. 

The constant $F_Q$ that appears in the couplings is the PNGB
decay constant. Its value is model-dependent. In this work we consider
three values for $F_Q$: $F_Q=125$~GeV for the one family technicolor
model, $F_Q=80$~GeV for top-color assisted technicolor and $F_Q=40$~GeV 
for multiscale technicolor.

Some typical values of the technieta partial and total widths are shown in 
Table \ref{tab:widths} for
$\alpha_s=0.119$ and $M_{\eta_T}=250$~GeV.

\section{Signal and Background Rates}

 With the  couplings discussed in the previous
 section we can show that the partonic cross section for the process
 $gg\rightarrow \eta_T \rightarrow \gamma g$ can be written as:

 \begin{equation}
   \label{eq:partcs}
   \hat{\sigma}=\frac{5\hat{s}^3\pi^3}{384}\left(\frac{N_T e g_s}{12\sqrt{2}\pi
  F_Q}\right)^2\left(\frac{N_T \alpha_s}{\sqrt{2}\pi
  F_Q}\right)^2 \frac{1}{(\hat{s}-M_{\eta_T}^2)^2+\Gamma_{\eta_T}^2M_{\eta_T}^2 }
 \end{equation}

We wrote a Fortran code in order to convolute the above partonic cross
section with the CTEQ4M\cite{CTEQ} partonic
distribution functions (with $Q^2=M_{\eta_T}^2$). Because the
technieta coupling with quarks is proportional to the quark mass, we
neglect the technieta production via $q\bar{q}\eta_T$ interaction. In
the case of gluon fusion we only take into account the $s$-channel contribution,
which is dominant at the resonance. It
must be noted that gauge invariance is preserved due to  the Levi-Civita
tensor present in equation (\ref{etagfoton}). Table \ref{tabsinal}
shows the cross section (in pb) calculated for different values of
$M_{\eta_T}$ and $F_Q$ at $\sqrt{s}=2000$~GeV with a cut in the
transverse photon and jet momentum $p_{T\gamma,j}>10$~GeV. 
These values for the cross section agree, 
with a precision of one
percent, with a narrow width approximation. 
The cross section becomes sizeable for low values $M_{\eta_T}$ and $F_Q$,
being of the order of a picobarn. 
However, the cross section  for the background 
$p\bar{p}\rightarrow\gamma g$ and
$p\bar{p}\rightarrow\gamma q$ processes is $\sigma_{\mbox{back}}=2.14 \times 10^4$
pb, which is a factor of $10^4$
larger than the signal. This situation clearly shows that a detailed
kinematical analysis is necessary to work out the strategy to suppress 
the background 
as strongly as possible in order to extract the signal.

\section{Complete Simulation of Signal and Background}

In order to perform a complete signal and background simulation we use 
the PYTHIA 5.7
\cite{PYTHIA} generator. Effects of jet fragmentation, initial and
final state radiation (ISR+FSR) as well as smearing of
the  jet and the photon energies have been taken into account. Since the 
process $\eta_T\to
\gamma g$ was absent in PYTHIA we created a generator for  $gg\to\eta_T
\to\gamma g$ process and linked it to PYTHIA as an external user process.

In our simulation we have used CTEQ4M structure function and 
have chosen $Q^2=M_{\eta_T}^2$ for the signal.

\begin{figure}
\vspace*{-1cm}
\begin{picture}(400,640)(0,0)
\put(0,  410){\epsfxsize=0.8\hsize \epsffile{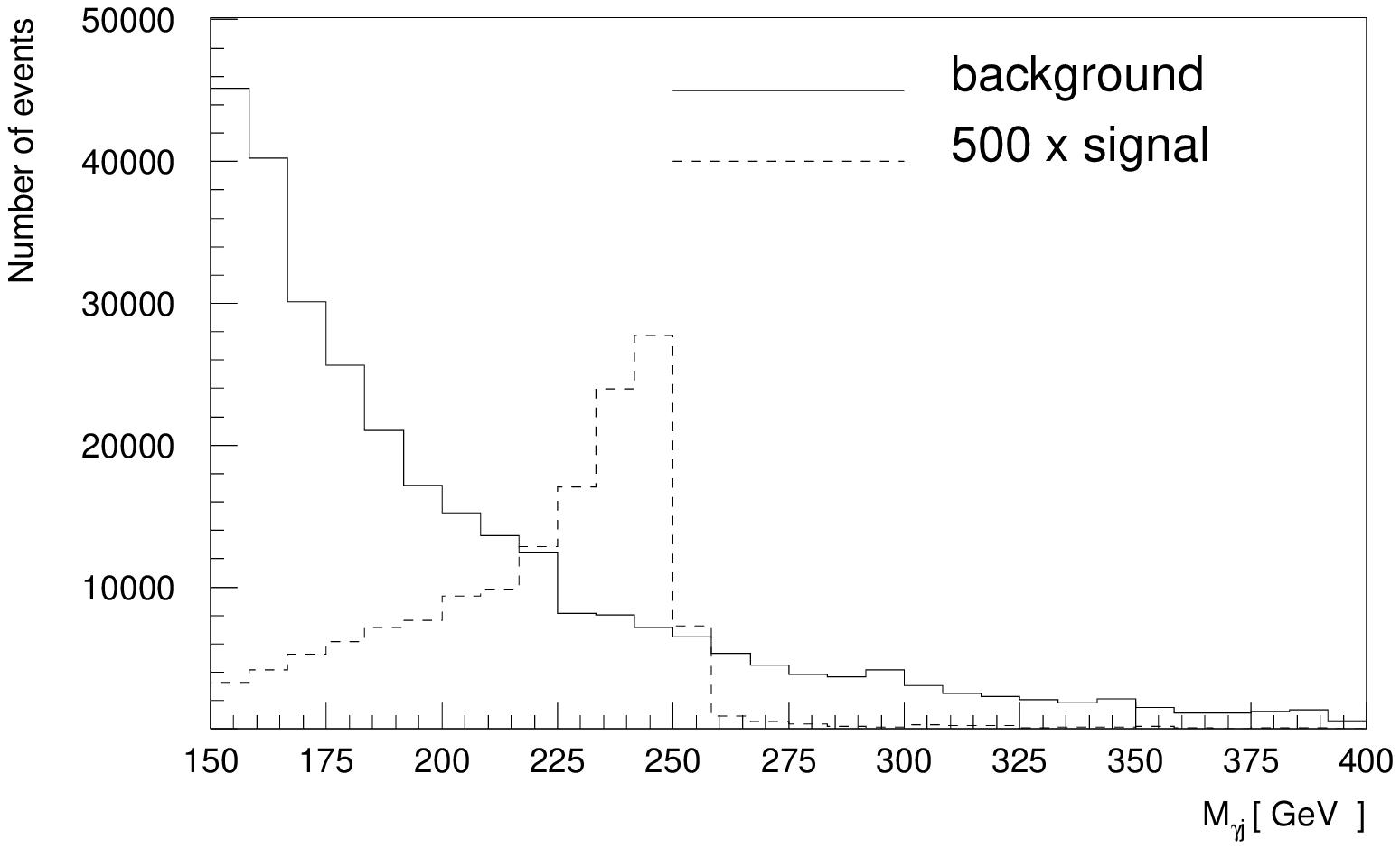}   }
\put(0,  220){\epsfxsize=0.5\hsize \epsffile{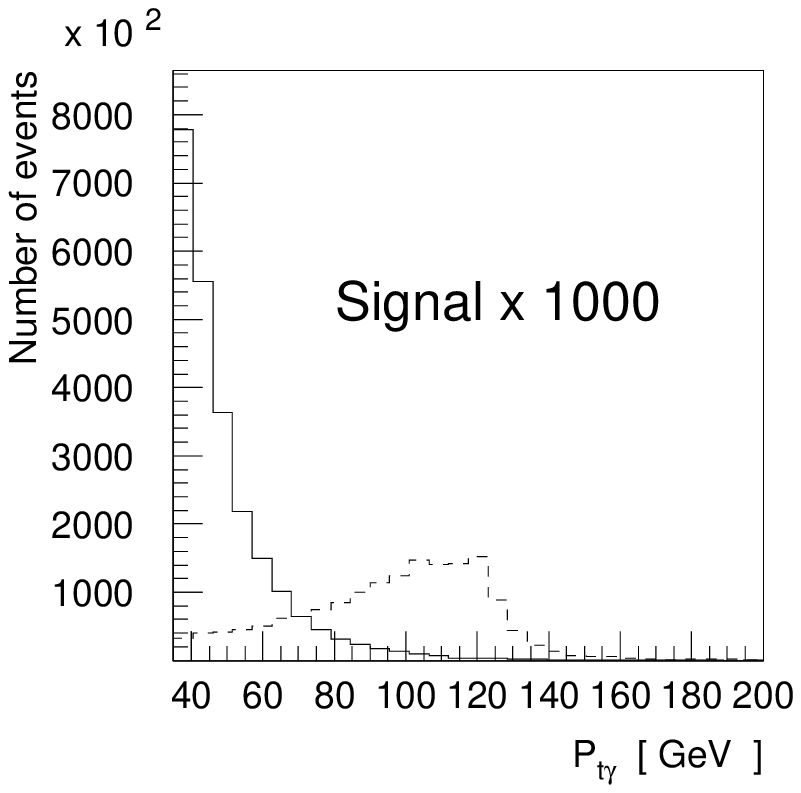}    }
\put(220,220){\epsfxsize=0.5\hsize \epsffile{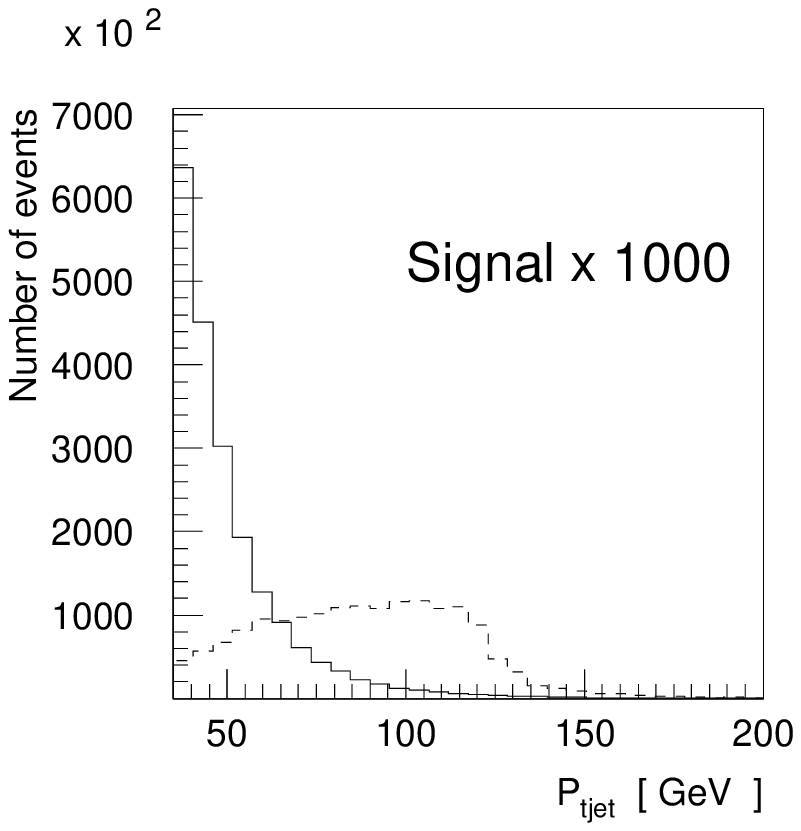}    }
\put(0,   30){\epsfxsize=0.5\hsize \epsffile{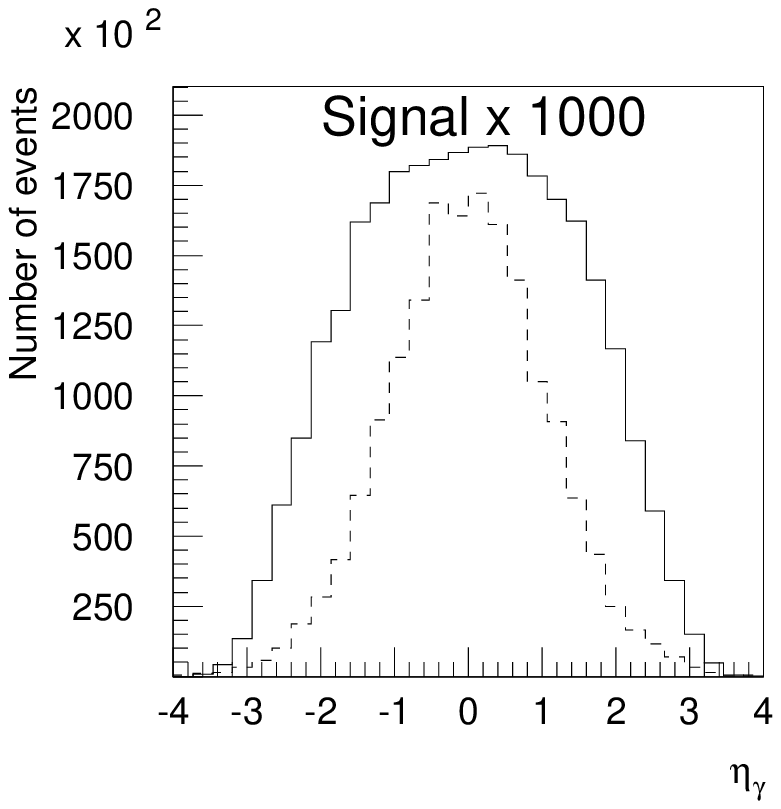}   }
\put(220, 30){\epsfxsize=0.5\hsize \epsffile{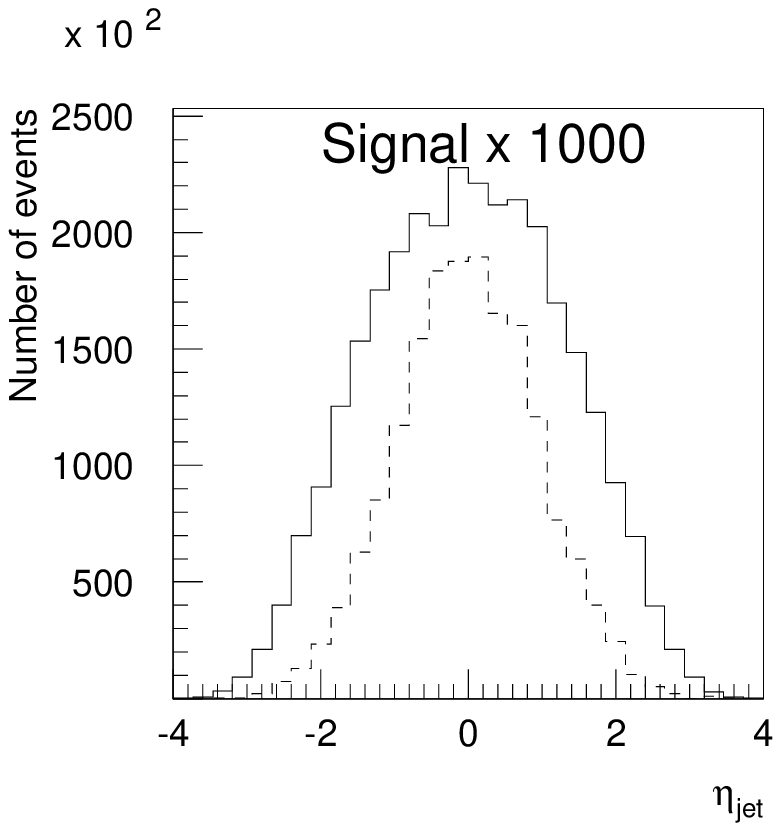}   }
\put(80,   600){a)}
\put(60,   390){b)}
\put(280,  390){c)}
\put(60,   200){d)}
\put(280,  200){e)}
\end{picture}
\vspace*{-1.7cm}
\caption{Distributions of invariant mass (a), photon tranverse
  momentun (b), jet tranverse momentun (c), photon rapidity (d) and
  jet rapidity (e) for signal (dashed line) and background (solid
  line) before any cut. We have assumed
  $M_{\eta_T}=250$ GeV and $F_Q=40$ GeV.}
\label{dist}
\end{figure}

In this framework we study, for both signal and background, 
distributions of the 
transverse photon momentum, transverse jet momentum, rapidity and invariant
mass in order to find the optimal kinematical cuts for signal subtraction. These
distributions are shown in Fig.~\ref{dist}. We can see that the ISR+FSR and
energy smearing effects make the mass distribution (Fig.~\ref{dist}(a)) quite
broad.  Notice the difference in $p_t$ distribution for
photons(Fig.~\ref{dist}(b)) and jets(Fig.~\ref{dist}(c)).  The fact that the
distribution for jets is wider than for photons is due to initial and final
state radiation. 

We have found the following optimal set of kinematical cuts:
\begin{eqnarray}
  \label{eq:ptcut}
 && p_{t\gamma \mbox{\small ,jet}}>\frac{M_{\eta_T}}{2}-40\mbox{ GeV}\\
 &&  \label{eq:invmasscut}
 M_{\eta_T}-\frac{M_{\eta_T}}{10}\leq M_{\gamma \mbox{\small jet}}\leq
 M_{\eta_T}+10\mbox{ GeV}
\end{eqnarray}
To take into account the detector pseudorapidity coverage  we have chosen the
following cuts for $\eta_\gamma$ and $\eta_{jet}$:
\begin{eqnarray}
   \label{eq:rapcut}
  |\eta_{\gamma}|\leq 1.5, \ \ |\eta_{jet}|<3
\end{eqnarray}

\begin{figure}
\begin{picture}(400,250)(0,0)
\put(-30,     10){\epsfxsize=0.6\hsize\epsffile{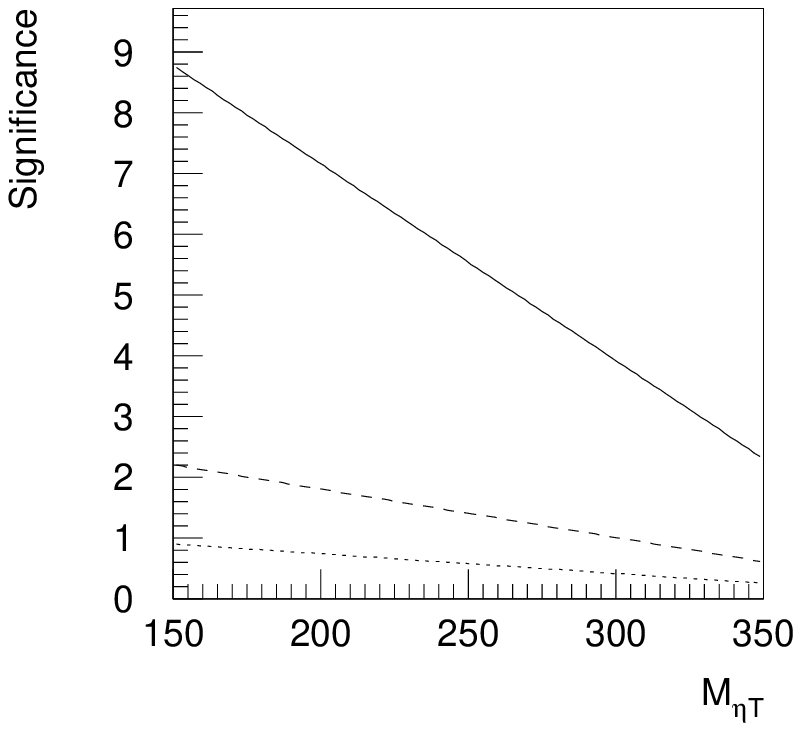}}
\put(220,   10){\epsfxsize=0.6\hsize\epsffile{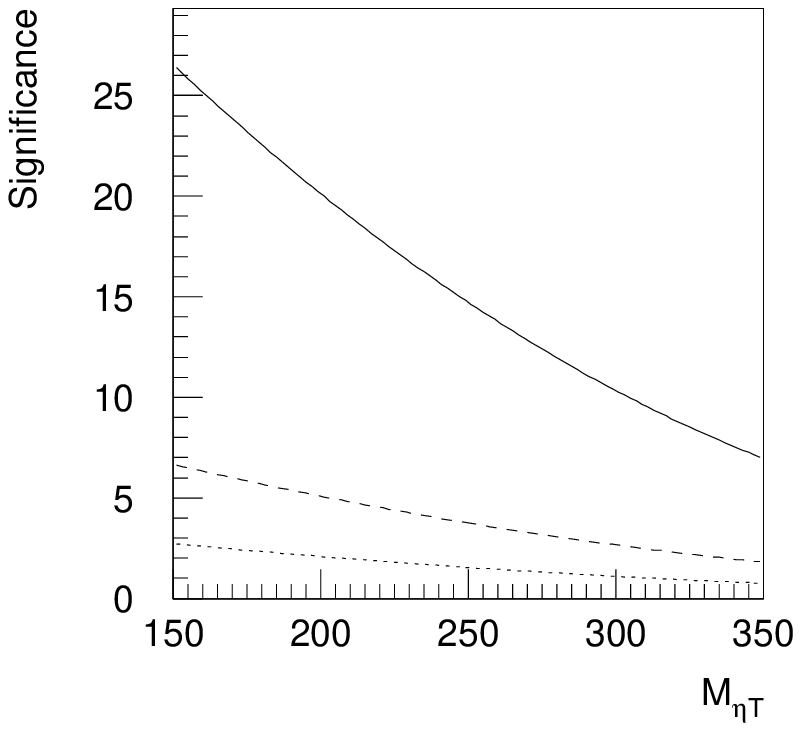}}
\put( 40, 220){a)}
\put(290, 220){b)}
\end{picture}
\vspace*{-0.5cm}
\caption{Significance as a function of $M_{\eta_T}$ for multiscale
  technicolor (solid line), top-color assisted technicolor (dashed
  line) and one family technicolor (dotted line), based on results
  obtained with (a) and without (b) taking into account ISR+FSR and
  energy smearing effects.}
\label{fig:signif}
\end{figure}

Table \ref{tab:signal_after_cuts} shows the signal and background cross sections
after those cuts have been applied. It is interesting to look at the values of
the significance which is written as  $\frac{{\cal
L}\sigma_{\mbox{signal}}}{\sqrt{{\cal L}\sigma_{\mbox{back}}}}$ and
characterizes the statistical deviation of the number of the observed events
from the predicted background. The significance as a function of the 
$M_{\eta_T}$
for different technicolor  models is shown in  (Fig.\ref{fig:signif}(a),
where we have assumed a luminosity of  ${\cal L}= 2000$ pb$^{-1}$ for the
Tevatron Run II.  For  multiscale technicolor ($F_Q=40$~GeV), the significance
is above the $2\sigma$ 95\%~CL exclusion limit for technieta mass less
than 350~GeV while for a $5\sigma$ discovery criteria one obtains
$M_{\eta_T}>266$~GeV mass limit. For the top-color assisted technicolor model
($F_Q=80$~GeV) one can establish only a 95\%~CL exclusion limit 
$M_{\eta_T}>175$~GeV. For the  one family technicolor model the significance 
is too
small  to establish any limits on $M_{\eta_T}$.

In our study we compared results based on PYTHIA simulation
and results obtained using  MADGRAPH~\cite{MAD} and HELAS\cite{HELAS} 
without taking into account ISR+FSR and  the energy 
smearing effects. The corresponding significance for this case is shown 
in Fig.\ref{fig:signif}(b). One can see that for this ideal case
respective values of significance is about 2.5 times higher than
in the case when we model the realistic situation using PYTHIA.
The differences between  Fig.\ref{fig:signif} (a) and (b)
clearly shows the importance of the complete simulation of the signal and 
background in order to obtain realistic results.

Finally it is worth pointing out that the study of the $b\bar{b}$ signature 
would lead to similar bounds on the $\eta_T$ mass. This is because the signal
for $b\bar{b}$ final state will be roughly increased by a factor 10  (see
Table~\ref{tab:widths}) but the background will be about two orders higher 
than that for $\gamma+jet$ signature. This would lead to the same values of
the significance as for $\gamma+jet$ final state. However, one should take into
account also the efficiency of b-tagging which will decrease the significance. 

\section*{Conclusions}

We have studied the potential of the upgraded Tevatron collider for the $\eta_T$
search with  $\eta_T \rightarrow \gamma + g$ decay signature and mass below the
$t\bar{t}$ threshold.  Results have been obtained for  the one family model,
top-color assisted technicolor and multiscale technicolor.

We found that for multiscale technicolor model, Tevatron can exclude 
$M_{\eta_T}$ up to $350$~GeV at $95$\%CL, while the $5 \sigma$ discovery limit 
for $\eta_T$ is 
$266$~GeV. For the
top-color assisted technicolor model one can only put a $95$\%CL lower limit on
$\eta_T$ mass equal to $175$~GeV while for one family technicolor model
the significance is too small  to establish any limit at all. 
Study of $b\bar{b}$
final state signature is not expected to give better limits on the  $\eta_T$
mass.

We have performed a complete simulation of the signal and background  and have
shown the importance of taking into account the effects of jet fragmentation, 
initial
and final state radiation, as well as smearing of the
jet and the photon energies.

\acknowledgments
This work was supported by Conselho
Nacional de Desenvolvimento Cient\'{\i}fico e Tecnol\'ogico
(CNPq), by Funda\c{c}\~ao de Amparo \`a Pesquisa do Estado de
S\~ao Paulo (FAPESP), and by Programa de Apoio a N\'ucleos de
Excel\^encia (PRONEX).

\begin{table}[htbp]
  \begin{center}
  \begin{tabular}{|c|c|c|c|}   \hline
\multicolumn{1}{|c|}{$\Gamma\setminus F_Q$}
&\multicolumn{1}{c|}{$40$~GeV} 
&\multicolumn{1}{c|}{$80$~GeV}&\multicolumn{1}{c|}{$125$~GeV}
\\ \hline
$\eta_T \rightarrow ag $&$0.008$ &$0.002$ &$8\times10^{-4}$ \\ \hline
$\eta_T \rightarrow gg $&$0.929$ &$0.232$ &$0.095$ \\ \hline
$\eta_T \rightarrow b\bar{b}$&$0.078$ &$0.019$ &$0.008$ \\ \hline
$\eta_T \rightarrow \mbox{all} $&$1.015$ &$0.254$ &$0.104$ \\\hline
\end{tabular}
    \caption{Partial widths (in GeV) for $M_{\eta_T}=250$~GeV.}
    \label{tab:widths}
  \end{center}
\end{table}

\begin{table}[htbp]
\begin{center}
\begin{tabular}{|c|c|c|c|}   \hline
\multicolumn{1}{|c|}{$M_{\eta_T}\setminus F_Q$}
&\multicolumn{1}{c|}{$40$~GeV} 
&\multicolumn{1}{c|}{$80$~GeV}&\multicolumn{1}{c|}{$125$~GeV}
\\ \hline
$150$~GeV &6.00  &1.50& 0.62   \\ \hline 
$200$~GeV & 2.58 &0.65& 0.27  \\ \hline
$250$~GeV &1.17 &0.29 &0.12   \\ \hline
$300$~GeV &0.55 &0.14 &0.06    \\ \hline
$340$~GeV &0.31 &0.08 &0.03    \\ \hline
\end{tabular}
\caption{Total cross section in pb for the process $p\bar{p}\rightarrow\eta_T
\rightarrow g \gamma$ with $\sqrt{s}=2000$~GeV.  }
\label{tabsinal}
\end{center}
\end{table}

\begin{table}[htbp]
\begin{center}
\begin{tabular}{|c|c|c|c|c|}   \hline
\multicolumn{1}{|c|}{$M_{\eta_T}\setminus F_Q$}
&\multicolumn{1}{c|}{$40$~GeV} 
&\multicolumn{1}{c|}{$80$~GeV}&\multicolumn{1}{c|}{$125$~GeV}
&\multicolumn{1}{c|}{Background}
\\ \hline
$150$~GeV &1.53&0.38&0.16 &60.72    \\ \hline 
$200$~GeV &0.63&0.16&0.06 &15.51   \\ \hline
$250$~GeV &0.26&0.07&0.03 &4.53  \\ \hline
$300$~GeV &0.11&0.03&0.01 &1.72   \\ \hline
$340$~GeV &0.06&0.01&$6\times10^{-3}$ &0.94   \\ \hline
\end{tabular}
\caption{Signal and background cross sections in pb for $\sqrt{s}=2000$~GeV 
after cuts. }
\label{tab:signal_after_cuts}
\end{center}
\end{table}

\begin{figure}
\begin{picture}(400,300)(0,0)
\put(0,   10){\epsfxsize=\hsize \epsffile{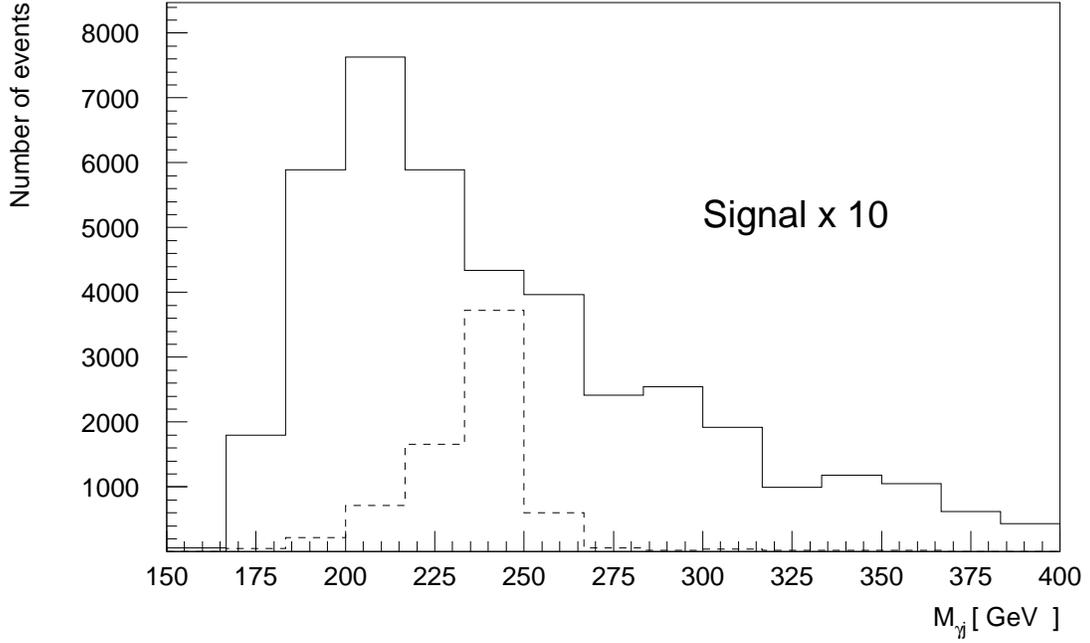}}
\end{picture}
\vspace*{-0.5cm}
\caption{Invariant mass distribution for signal (dashed line) and
  background (solid line) after cuts,
  $M_{\eta_T}=250$ GeV and $F_Q=40$ GeV.}
\label{}
\end{figure}

\begin{references}

\bibitem{SUSY}Y.~Golfand and E.~Likhtman, {\em JETP Lett.} {\bf 13},
  (1971)323; D.~Volkov and V.~Akulov, \PLB{46}{73}{109}; J.~Wess and B
  Zumino, \NPB{70}{74}{39}.


\bibitem{extradim}N.~Arkani-Hamed, S.~Dimopoulos and
  G.~Dvali,\PLB{429}{98} 263.~

\bibitem{Weinberg79}S.~Weinberg, \PRD{19}{79}{1277}.

\bibitem{Suskin79}L.~Susskind, \PRD{20}{79}{2619}.

\bibitem{chivros}R.~S.~Chivukula, R.~Rosenfeld, E.~H.~Simmons and
  J.~Terning, in {\em Electroweak Symmetry Breaking and Beyond the
  Standard Model}, edited by T.~Barklow, S.~Dawson, H.~E.~Haber, and
  J.~Siegrist,  World Scientific.

\bibitem{HCI}E.~Eichten and K.~Lane,
  Proceeding of the 1996 DPF/DPB Summer Study on New Directions for
  High Energy Physics (Snowmass 96); K.~Lane, \PLB{357}{95}{624};
  S.~Mrenna and J.~Womersley,
  \PLB{451}{99} 155; K.~Lane, hep-ph/9903372.

\bibitem{Lubicz}V.~Lubicz and P.~Santorelli,\NPB{460}{96} 3;
  G.~Rupak and E.~H.~Simmons, \PLB{362}{95}{155}.

\bibitem{NLC}W.~Skiba, \NPB{470}{96} 84; R.~Rosenfeld and
  A.~Zerwekh, \PLB{418}{98} 329.


\bibitem{TFMC}K.~Lane, talk presented at the Workshop on Physics at
  the First Muon Collider and at the Front End of a Muon Collider,
  Fermilab, November 6-9, 1997, hep-ph/98011385; E.~Eichten, K.~Lane
  and J.~Womerslay, \PRL{80}{98} 5489.

\bibitem{WTC}B.~Holdom,  {\em Phys.~Rev} {\bf D24} (1981) 1441;
B.~Holdom, {\em Phys.~Lett.} {\bf B150} (1985) 301; K.~Yamawaki,
M.~Bando, and K.~Matumoto, {\em Phys.~Rev.~Lett.}{\bf 56} (1986) 1335;
T.~Appelquist, D.~Karabali, and L.C.R.~Wijewardhana, {\em
  Phys.~Rev.Lett.} {\em 57} (1986) 957; T.~Appelquist and
L.C.R.~Wijewardhana, {\em Phys.~Rev} {\bf D35} (1987) 774;
T.~Appelquist and L.C.R.~Wijewardhana, {\em Phys.~Rev} {\bf D36}
(1987) 568.

\bibitem{Klimits}Z.~Xiao, L.~L\"u, H.~Guo and G.~Lu, hep-ph/9903347.

\bibitem{MTC}K.~Lane and E.~Eichten, {\em Phys.~Lett.} {\bf B222}
(1989) 274.

\bibitem{ttbar1}T.~Appelquist and G.~Triantaphyllou,
  \PRL{69}{92} 2750.

\bibitem{OFTC}E.~Farhi and L.~Susskind, {\em Phys. Rev.} {\bf D20} (1979)
3404.


\bibitem{ttbar2}E.~Eichten and K.~Lane,\PLB{327}{94} 129.

\bibitem{eta1}F.~Hayot and O.~Napoly, \ZPC{7}{81} 229.


\bibitem{TC2}C.~T.~Hill,\PLB{345}{95}{483}.

\bibitem{ABJ}S.~L.~Adler, {\em Phys.~Rev.} {\bf 177} (1969) 2426;
  J.~S.~Bell and R.~Jackiw, \NC{60}{69}{47}.

\bibitem{Ellis}J.~Ellis, M.~K.~Gaillard, D.~V.~Nanopoulos and
  P.~Sikivie, \NPB{182}{81} 529.

\bibitem{CTEQ}H.~L.~Lai et al., CTEQ Collab., \PRD{55}{97}{1280}.

\bibitem{MAD}T.~Stelzer and W.~F.~Long, {\em Comput. Phys. Commun.}
{\bf 81} (1994) 357.

\bibitem{HELAS}H.~Murayama, I.~Watanabe and K.~Hagiwara, KEK report 
No.~91-11, unpublished.

\bibitem{PYTHIA}T.~Sj\"ostrand, {\em Comput. Phys. Commun.} {\bf 82} (1994) 74.
\end{references}
\end{document}